\documentclass[12pt,preprint]{aastex}
\begin{document}

\title{Simulating a White Dwarf-dominated Galactic Halo}
\author{Chris B. Brook, Daisuke Kawata \& Brad K. Gibson\\
        Centre for Astrophysics \& Supercomputing, 
        Swinburne University, Mail \#31, P.O. Box 218,
	Hawthorn, Victoria, 3122, Australia}


\begin{abstract}
Observational evidence has suggested the possibility of a Galactic halo
which is dominated by white dwarfs (WDs).  While debate continues concerning
the interpretation of this evidence, it is clear that an initial mass
function (IMF) biased heavily 
toward WD precursors (1$<$m/M$_\odot$$<$8), at least in the
early Universe, would be necessary in generating such a halo.  Within the
framework of homogeneous, closed-box models of Galaxy formation, such
biased IMFs lead to an unavoidable overproduction of carbon and nitrogen
relative to oxygen (as measured against the abundance patterns in the oldest
stars of the Milky Way).  
Using a three-dimensional Tree N-body smoothed particle hydrodynamics code,
we study the dynamics and chemical evolution of a galaxy with different
IMFs.  Both invariant and metallicity-dependent IMFs are considered.
Our variable IMF model invokes a WD-precursor-dominated IMF for metallicities
less than 5\% solar (primarily the Galactic halo), 
and the canonical Salpeter IMF otherwise (primarily the disk).  Halo WD
density distributions and C,N/O abundance patterns are presented.  While
Galactic haloes comprised of $\sim $5\% (by mass) of WDs are not supported by our simulations, mass fractions of $\sim$1-2\% cannot be ruled out.
This conclusion is consistent with the present-day observational constraints.
\end{abstract}

\clearpage
\section{Introduction}\label{densection}

Evidence gathered by microlensing surveys toward the Large Magellanic
Cloud suggest that a substantial fraction of the Milky Way's dark matter
halo is comprised of $\sim$0.5\,M$_\odot$ objects - the MACHO Team claim
20\% of the halo's dynamical mass may be tied up in these half-solar
mass constituents (Alcock et~al. 2000), while the EROS Team favour an
upper limit of 30\% (Lasserre et~al. 2000).
Constraints set by Hubble Space Telescope
red star counts (Flynn et~al. 1996) rule out large numbers of low-mass hydrogen
burning stars in the halo.  By process of elimination, white dwarfs (WDs) 
represent one potential baryonic candidate for these 
microlensing events.  This would correspond to of order
2$\times$10$^{11}$ WDs out to a distance of 60 kpc in the Galactic halo (see Zritsky 1999).

The apparent identification of 2-5 high-proper motion WDs in
the Hubble Deep Field-North supported the WD-dominated
halo hypothesis (Ibata et~al. 1999),
as did the colour analysis of the Hubble Deep Field-South (Mendez et~al.
2000) - the Ibata et~al. claim though has since been retracted (Richer 2001).
Ground-based searches for nearby high-proper motion WDs initially
suggested a WD halo mass fraction of 10\% (Ibata et~al. 2000), although 
Flynn et~al. (2001) favoured $\sim $2\%.  Based upon dynamical
arguments and Galactic winds driven by a putative
population of WD progenitors, Zhao (2002) suggest a fraction
$<$4\%.  Regardless, Oppenheimer et~al. (2001, hereafter OHDHS) have 
recently revived the
WD-dominated halo picture with their claim that $>$2\% of the halo
is made up of WDs.  While criticism of the OHDHS analysis abounds (Gibson
\& Flynn 2001; Reid et~al. 2001; Koopmans \& Blandford 2002; Reyl\'e et~al.
2001; Flynn et~al. 2002; Torres et~al. 2002), 
their result\footnote{In addition to that of 
Nelson et~al. (2002), who claim a 7\% WD dark halo based upon 24 candidate 
high-proper motion objects in the Groth Strip, and Mendez (2002), who
claims (based upon one apparent high-proper motion white dwarf seen towards
NGC~6397) that most of the dark matter in the solar vincinity can be accounted
for by halo and thick-disk white dwarfs.} (if confirmed) may yet 
be made consistent with the MACHO and EROS Team\footnote{The EROS-2
white dwarf proper motion survey find that the halo white 
dwarf fraction cannot exceed 5\% (95\% c.l.).} results.

WDs represent the end point in the life cycle of stars whose initial
masses are in the approximate range $1-8$\,M$_\odot$.  Populating
a WD-dominated Galactic halo, in conjunction with a
standard stellar initial mass function (IMF), would lead to a 
corresponding increase in the production of Type~II supernovae (SNeII) and
the rapid overproduction of heavy elements (Gibson \& Mould 1997,
hereafter GM97).  In addition, the expected number of low-mass red
dwarfs would violate the deep halo star counts  of Flynn et.~al. (1996). 
These high- and low-mass ``constraints'' imply that a standard IMF
cannot be employed to populate a putative WD-dominated Galactic halo.
Instead, the IMF would have had to have been comprised almost
exclusively of stars in the range 1$< $m/M$_\odot$$<$8 - such WD-heavy
IMFs have been discussed in the literature by
Chabrier et~al. (1996 - hereafter C96) and Adams \& Laughlin (1996).

Besides the requirement that any WD-heavy IMF be restricted to
1$<$m/M$_\odot$$<$8, there is an additional constraint provided
by the observed abundance patterns in Population~II halo stars.
GM97 found that a WD-dominated IMF (wdIMF) inevitably leads to the
overproduction of carbon and nitrogen relative to oxygen - \it by more
than an order of magnitude \rm - in comparison with that seen in the oldest
stars of the Milky Way's halo.  This result is a natural consequence of 
the life cycle of low- and intermediate-mass stars, stars which are
responsible for synthesising the majority of the carbon and nitrogen in
the Universe.  Fields et~al. (2000) confirmed the analysis of GM97,
concluding that the halo carbon and nitrogen arguent provides the
strictest constraint to any WD-dominated halo scenario.

While there is no compelling direct evidence for variable IMFs,
\it at the present time \rm (e.g. Gilmore 2001), several theoretical
studies suggest that the IMF of the first generation of stars
(so-called Population~III) might be quite different from that of the
canonical Salpeter IMF (sIMF). Direct simulation attempts to form Pop~III stars indicate an IMF biased toward super massive stars (M $> 100$ M$_\odot$) rather than white dwarf progenitors (Abel, Bryan \& Norman 1999, 2002). It is the lack of metals, which cause cloud fragmentation of gas in  star forming regions, that results in such a top heavy IMF (see also Bromm et.~al 2001). Other studies also indicate a different IMF for Pop~III stars, e.g., Yoshii \& Saio (1987) and 
Nakamura \& Umemura (2001), which suggested a bimodal IMF making large numbers of super massive stars around 100 M$_\odot$ and a futher peak in the intermediate, white dwarf progenitor mass range. It is possible that the stars from the high mass peak of such an IMF would not contribute to chemical enrichment due to the formation of black holes, thus making such an IMF effectively favour white dwarf progenitors. This Pop~III IMF is expected to apply to the
first generation of halo stars - the generation which would manifest itself
\it now \rm in the population of putative high-proper motion stellar halo WDs. 
Such studies remain, however, highly uncertain.   Our motivation for invoking the wdIMF remains the MACHO observations and the necessity for invoking such an IMF in order for these observations to be explained by a large population of white dwarf stars populating the halo.  
In addition to the wdIMF and sIMF possibilities, in what follows
we will also consider a variable IMF in which stars formed from the low (high)  metallicity gas are distributed following the adoption of
the wdIMF (sIMF).  One can anticipate that while the
wdIMF phase will invariably lead to high C/O and N/O ratios (as discussed
by GM97), the subsequent 
sIMF phase can ameliorate this effect. This variable IMF scenario was not
considered by GM97.   

The currently favoured hierarchical clustering scenario of galaxy
formation postulates that star formation progresses through the on-going
accretion of galactic building blocks. Within this framework,
highly inhomogeneous chemical evolution occurs; the assumption of
homogeneous closed-box models for the Milky Way's evolution (as was 
admittedly adopted by GM97) are no longer suitable.  The early phases of 
galaxy formation demand a chemo-dynamical approach, one which
we adopt here explicitly within the context of the WD-dominated IMF
hypothesis.  In what follows, we examine the ability of metallicity-dependent
variable IMFs to resolve the apparent contradiction between a 
significantly WD-heavy Galactic halo and the C/O and N/O abundance patterns
of halo stars.  To achieve this, we perform
numerical simulations which calculate self-consistently 
the chemical and dynamical evolution of a Milky Way-like galaxy.

In Section~2, we describe our numerical methods, concentrating upon
the code itself and our treatments of star formation, IMFs, and energy
feedback to the interstellar medium (ISM).  Section~3 presents the
results of our simulations in terms of resulting white dwarf halo
densities, and Galactic chemical evolution of carbon, nitrogen, and
oxygen.  The implications for a WD-dominated Galactic dark matter halo
are described in Section~4.

\section{Method}

\subsection{The Code}

Our simulations were performed using a modified version of the software
package described by Kawata (1999, 2001).  An overview of the code is provided 
here along with the relevant modifications made for the current study - 
the reader is referred to Kawata (1999, 2001) for all other details.
Our code is based upon the Hernquist \& Katz 1989 and Katz,
Weinberg \& Hernquist 1996 {\tt TreeSPH}, combining
a tree algorithm (Barnes \& Hut 1986) with smoothed particle hydrodynamics 
(SPH) (Lucy 1977; Gingold \& Monaghan 1977) for the computation of 
gravitational forces and numerical
hydrodynamics respectively.  This three-dimensional code is fully
Lagrangian, with individual smoothing lengths and time steps making it
highly adaptive in space and time.  It includes a self-consistent
treatment of physical processes governing galaxy formation, including
self-gravity, hydrodynamics, radiative cooling, star formation, supernovae
feedback, and metal enrichment.
We employ metallicity-dependent gas cooling rates derived from
MAPPINGS~III (Sutherland \& Dopita 1993); their implementation is
described by Kawata \& Gibson (2003).

\subsubsection{Star Formation}

Star formation prescriptions similar to those of
Katz (1992) and Katz, Weinberg \&
Hernquist (1996) were employed. Star formation occurs whenever
\begin{enumerate}
\item the gas density is greater than a critical density:\hfill\break
$\rho_{crit} = 2\times 10^{-25}$ g\,cm$^{-3}$ (Katz et~al. 1996), and
\item the gas velocity field is convergent: $\nabla \cdot v_i<0$.
\end{enumerate}
\noindent
Our adopted star formation law is written
\begin{equation}\label{sfr}
{d\rho_*\over dt} =-{d\rho_g\over dt}={c_*\rho_g\over \tau_g} 
\end{equation} 
\noindent
where ${d\rho_*\over dt}$ is the star formation rate (SFR).  
Equation~\ref{sfr} corresponds
to a Schmidt law in which the SFR is proportional to $\rho_g^{1.5}$. 
The dynamical timescale of the gas is given by
$\tau_g=\sqrt{(3\pi/16G\rho_g)}$; in
regions eligible for star formation, this dynamical timescale
is generally longer than the
cooling timescale.  Star formation efficiency is parameterised by the
dimensionless parameter $c_*$, which, after Kawata (1999, 2001), 
was set to $c_*=0.5$.

Equation~\ref{sfr} implies that the probability, p$_*$, with which one gas particle entirely transforms into a  star particle during a discrete time step, $\Delta t$, is
    p$_*=1-exp(-c_*\Delta t/t_g)$.	
This avoids an intolerably large number of star particles of different masses being formed. The newly created star particle behaves as a collisionless particle.

\begin{subsubsection}{Initial Mass Functions}

In our {\tt TreeSPH} simulations, ``stars''
are represented as particles with a typical mass of
order $10^5$\,M$_\odot$; the relative number distribution of stellar masses
within a ``particle'' is governed by the assumed initial mass function (IMF).
Fundamental to our modeling is the adoption of the Chabrier et~al. (1996)
white dwarf progenitor-dominated IMF (wdIMF).  We represent this IMF by a
truncated power-law of the form
\begin{equation}
\label{wdimf}
\Phi (m)\,{\rm d}m = dn/dm = Ae^{-\left({\bar{m}/m}\right)^{\beta}}\times
m^{-\alpha}\,{\rm d}m
\end{equation}
\noindent
for which we use $\bar{m}=2.7$, $\beta=2.2$, and $\alpha=5.75$.  The peak of 
this wdIMF occurs at $m\approx 2$\,M$_\odot$, favouring the production of
WD progenitors.

The canonical Salpeter (1955) IMF (sIMF) is used to describe
the local stellar mass function (by number) and is written
\begin{equation}\label{simf}
\Phi (m){\rm d}m = dn/dm = Am^{-(1+x)}\,{\rm d}m,
\end{equation}
with $x$=1.35.  The sIMF (solid line) and wdIMF (dashed line)
are shown in Figure~1.

The coefficient $A$ in equations~\ref{wdimf} and \ref{simf} is determined by 
normalising the respective IMFs over the mass range
0.1$\leq$$m$/M$_\odot$$\leq$40.0.  We will explore simulations in which 
the adopted IMF is invariant in space and time (and governed by either
the wdIMF or sIMF functions), as well as metallicity-dependent models 
(in which star formation in regions where the gas metallicity is less
than 5\% solar is governed by the wdIMF form, with the sIMF form applying
otherwise).

\begin{figure}
\label{fig1}
\plotone{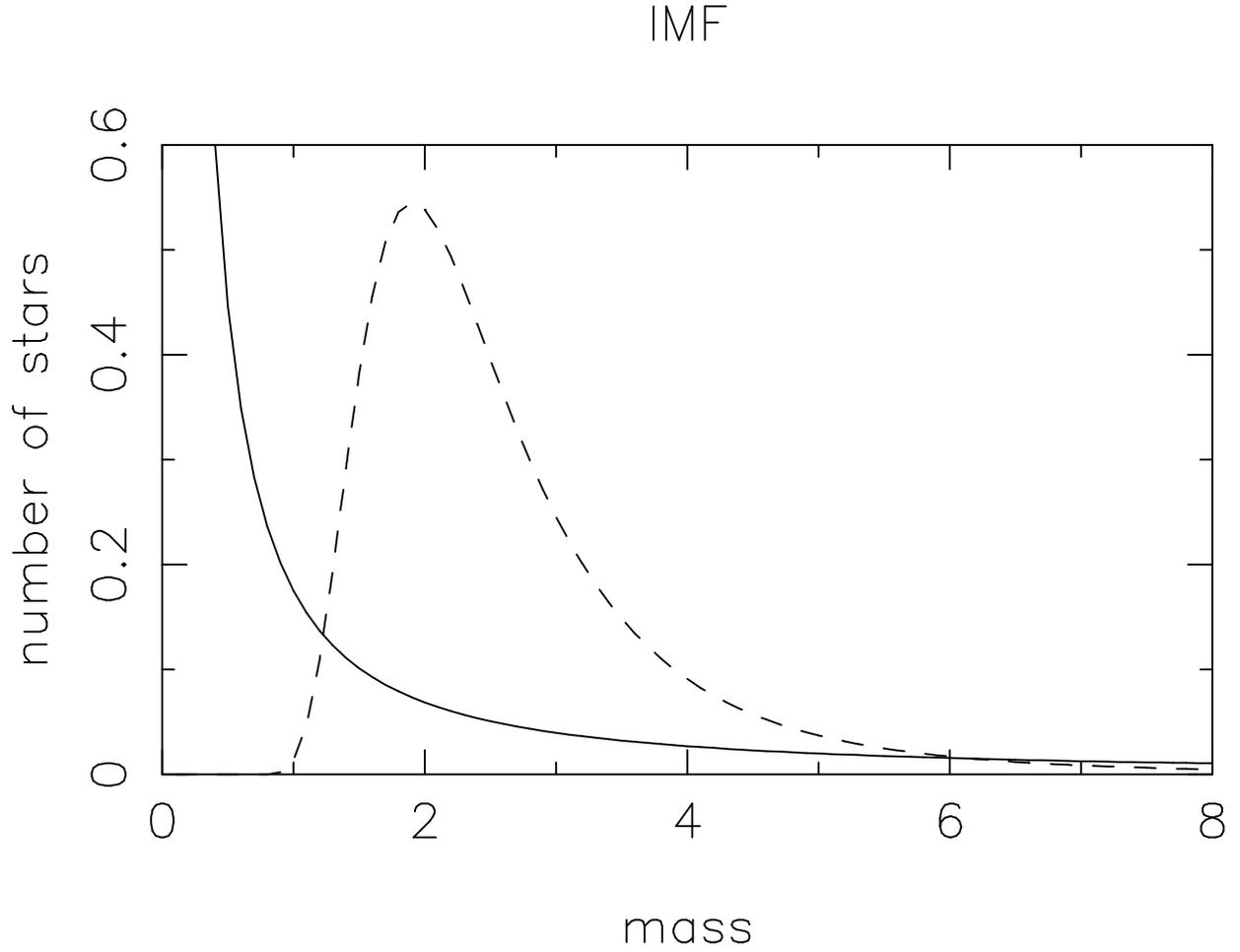}
\caption{Comparison of the Salpeter (1955) (solid
line) and Chabrier et~al. (1996) (dashed line) IMFs; the latter is dominated
by white dwarf progenitors.  Both IMFs have been normalised to unity over the 
mass range 0.1$\leq$$m$/M$_{\odot}$$\leq$40.0.}
\end{figure}

We will compare the results of our simulations using these very different
IMF assumptions in terms of their chemical evolution properties, and
contrast their inherent inhomogeneous nature with the original 
homogeneous models of GM97.  Each IMF assumption also leads to very
different outcomes for the halo and disk WD densities, an issue we
return to in Section~3.

\end{subsubsection}

\subsubsection{Feedback}

Type~II supernovae (SNeII) are assumed to be the end result for all stars 
of mass $m$$\geq$10\,M$_\odot$.
These massive stars explode within 
the simulation time step in which they are born, and so the
instantaneous recycling approximation (for the nucleosynthetic products
associated with these stars) applies.  SNeII release thermal and mechanical
energy to the surrounding interstellar medium, in addition to their
stellar yields; we assume
that each supernova releases $10^{50}$\,erg of thermal energy. 
For SNeII, the metallicity-dependant stellar yields of Woosley \& Weaver 
(1995) are adopted.

For low- and intermediate-mass stars, we use the stellar yields of
van~den~Hoek \& Groenwegen (1997).
The lifetimes of these stars are significantly
longer than the time steps of our simulations, and so the instantaneous 
recycling approximation is relaxed;  we assume that
recycling from intermediate mass stars occurs after a fixed time delay.  
Weighting by the IMF slope, over the range 1$\leq$$m$/M$_\odot$$\leq$8,
the typical mass is $\sim$2\,M$_\odot$ (for both the wdIMF and sIMF);
we therefore adopt a delay of 1~Gyr, corresponding to the main sequence
lifetime of a 2\,M$_\odot$ star (Schaller et~al. 1992).

For the current study, we ignore the effects of Type~Ia supernovae.
As noted by GM97, these supernovae do not produce significant amounts of
carbon, nitrogen, or oxygen (relative to SNeII), elements which form the
basis of the analysis which follows.

The mass, energy and heavy elements are smoothed over the neighbouring gas particles using the SPH smoothing algorithm. For example, in the timestep in which mass is released from a star particle due to a Type II supernova explosion, the increment of the mass of the $j$-th neighbour particle is
\begin{equation}\Delta M_{SN,j}={m_j\over \rho_{g,i}}M_{SN,i}W(r_{ij},h_{ij})\end{equation} where 
\begin{equation} \rho_{g,i}= \langle    \rho_g(x_i)\rangle =\sum_{j\neq i}m_j W(r_{ij},h_{ij})\end{equation}
and $ W(r_{ij},h_{ij})$ is an SPH kernal.

\begin{table*}\label{yields}
 \begin{minipage}{100mm}
  \caption{IMF-weighted stellar yields adopted for massive 
($m$$\geq$10\,M$_\odot$) stars.}
  \begin{tabular}{ccccc}
   & \multicolumn{4}{c}{Salpeter (1955) - sIMF} \\
Metallicity   & $0.0\leq$ Z/Z$_\odot<10^{-4}$ &  $10^{-4}\leq$ Z/Z$_\odot<10^{-1}$&   $10^{-1}\leq$ Z/Z$_\odot<1.0$&    $1.0\leq$ Z/Z$_\odot$\\
  Element       &           &                   &                  \\
 C  & $6.12\times 10^{-4}$  &  $8.46\times 10^{-4}$    &     $8.39\times 10^{-4}$ & $8.53\times 10^{-4}$ \\
 O  & $2.74\times 10^{-3}$ &   $5.99\times 10^{-3}$    &     $6.55\times 10^{-3}$   & $7.26\times 10^{-3}$ \\
 N  & $3.25\times 10^{-5}$  & $3.20\times 10^{-5}$     &      $6.09\times 10^{-5}$& $3.05\times 10^{-4}$  \\
 Ne & $2.08\times 10^{-4}$  &$7.32\times 10^{-4}$      &      $5.47\times 10^{-4}$& $6.92\times 10^{-4}$  \\
 Mg & $1.42\times 10^{-4}$  &$2.83\times 10^{-4}$      &      $3.17\times 10^{-4}$& $4.14\times 10^{-4}$ \\
 Si & $1.80\times 10^{-4}$  &$5.88\times 10^{-4}$      &      $6.06\times 10^{-4}$& $7.60\times 10^{-4}$ \\
 Fe & $4.78\times 10^{-4}$  &$4.76\times 10^{-4}$      &      $7.86\times 10^{-4}$& $7.05\times 10^{-4}$ \\
 $[$C/O$]$ & $-0.16$                &$-0.35$      &      $-0.40$& $-0.44$ \\
 $[$N/O$]$ & $-0.98$                &$-1.33$      &      $-1.09$& $-0.44$ \\
 & &  & & \\
   & \multicolumn{4}{c}{Chabrier et~al. (1996) - wdIMF} \\
 C  & $3.62\times 10^{-6}$  &  $5.49\times 10^{-6}$    &     $6.06\times 10^{-6}$ & $5.77\times 10^{-6}$ \\
 O  & $7.71\times 10^{-6}$ &   $1.47\times 10^{-5}$    &   $1.76\times 10^{-5}$   & $2.12\times 10^{-5}$ \\
 N  & $1.20\times 10^{-6}$  & $1.21\times 10^{-6}$     &      $1.47\times 10^{-6}$& $3.31\times 10^{-6}$  \\
 Ne & $8.66\times 10^{-7}$  &$6.14\times 10^{-6}$      &      $1.61\times 10^{-6}$& $3.07\times 10^{-5}$  \\
 Mg & $5.63\times 10^{-7}$  &$7.32\times 10^{-7}$      &      $8.30\times 10^{-7}$& $1.36\times 10^{-6}$ \\
 Si & $1.72\times 10^{-6}$  &$3.62\times 10^{-6}$      &      $2.92\times 10^{-6}$& $4.04\times 10^{-6}$ \\
 Fe & $6.00\times 10^{-6}$  &$3.94\times 10^{-6}$      &      $1.02\times 10^{-5}$& $6.41\times 10^{-6}$ \\
 $[$C/O$]$ & $0.167$                &$0.067$      &      $0.032$& $-0.070$ \\
 $[$N/O$]$ & $0.14$                &$-0.14$      &      $-0.13$& $0.13$ \\
\end{tabular}
\end{minipage}
\end{table*}

\begin{table*}\label{yields2}
\begin{minipage}{75mm}
  \caption{IMF-weighted stellar yields adopted for low- and intermediate-mass
($m$$\leq$8\,M$_\odot$) stars.}
  \begin{tabular}{ccccc}
&   & \multicolumn{2}{c}{Salpeter (1955) - sIMF}& \\
Metallicity   & $5\times 10^{-3}\leq$ Z/Z$_\odot<10^{-1}$ & $10^{-1}\leq$ Z/Z$_\odot<1.0$&     $1.0\leq$ Z/Z$_\odot$\\
  Element       &           &             &      &                  \\
 C  &$6.82 \times 10^{-4}$  &      $7.26\times 10^{-4}$ & $1.20\times 10^{-3}$ \\
 O  & $1.27 \times 10^{-5}$ &     $1.18\times 10^{-4}$   & $1.11\times 10^{-3}$ \\
 N  & $1.32 \times 10^{-3}$  &       $1.37\times 10^{-3}$& $1.21\times 10^{-3}$  \\
 Ne & $1.06 \times 10^{-5}$  &    $1.56\times 10^{-5}$& $2.36\times 10^{-5}$  \\
 Mg & $5.95 \times 10^{-7}$  &   $1.19\times 10^{-5}$& $1.26\times 10^{-4}$ \\
 Si &$6.46 \times 10^{-7}$  &      $1.29\times 10^{-5}$& $1.36\times 10^{-4}$ \\
 Fe &  $1.07 \times 10^{-6}$  &      $2.14\times 10^{-5}$& $2.26\times 10^{-4}$ \\
 $[$C/O$]$ & $2.22$      &      $1.28$& $0.529$ \\
 $[$N/O$]$ &$2.95$ &     $2.00$& $0.97$ \\
 & &  &  \\
&   & \multicolumn{2}{c}{Chabrier et~al. (1996) - wdIMF} \\
 C  &$3.96 \times 10^{-3}$     &     $3.10\times 10^{-3}$ & $3.96\times 10^{-3}$ \\
 O  &$7.67 \times 10^{-5}$     &   $4.54\times 10^{-4}$   & $4.31\times 10^{-3}$ \\
 N  & $2.42 \times 10^{-3}$    &      $1.43\times 10^{-3}$& $1.74\times 10^{-3}$  \\
 Ne & $3.32 \times 10^{-5}$   &      $5.70\times 10^{-5}$& $8.60\times 10^{-5}$  \\
 Mg &  $9.81 \times 10^{-7}$  &      $4.47\times 10^{-5}$& $4.70\times 10^{-4}$ \\
 Si & $1.07 \times 10^{-6}$   &      $4.85\times 10^{-5}$& $5.11\times 10^{-4}$ \\
 Fe &  $1.76 \times 10^{-6}$  &      $8.05\times 10^{-5}$& $8.48\times 10^{-4}$ \\
 $[$C/O$]$ &  $2.21$    &      $1.32$& $0.46$ \\
 $[$N/O$]$ &  $2.44$  &      $1.43$& $0.54$ \\
\end{tabular}
\end{minipage}
\end{table*}

Tables~1 (SNeII) and 2 (low- and intermediate-mass
stars) list the IMF-weighted stellar
yields (in solar masses ejected per solar mass of stars formed) adopted in
our analysis; in both tables, the first block represents the sIMF 
yields and the second block, the wdIMF.  Four (three) different metallicities
are highlighted for massive (intermediate mass) stars.

The overproduction of carbon and nitrogen with respect to oxygen,
for IMFs dominated by white dwarf precursors, can be best appreciated by
referral to Figure~2.
Shown here are the adopted stellar yields' 
[C,N/O]\footnote{[C/O]$\equiv$$\log$(C/O)$-$$\log$(C/O)$_\odot$} 
as a function of initial
mass; both [C/O] and [N/O] are significantly in excess of solar for
$m$$\leq$8\,M$_\odot$.  In comparison, the oldest stars of the Galactic
halo show [C/O]$\simeq$[N/O]$\simeq$$-$0.5 (Timmes, Woosley \& Weaver 1995). 
We can anticipate that an
IMF in which a large number of these intermediate mass stars are
present will result in an over-abundance of carbon and nitrogen compared
to solar abundances.

\begin{figure*}
\label{fig2}
\plotone{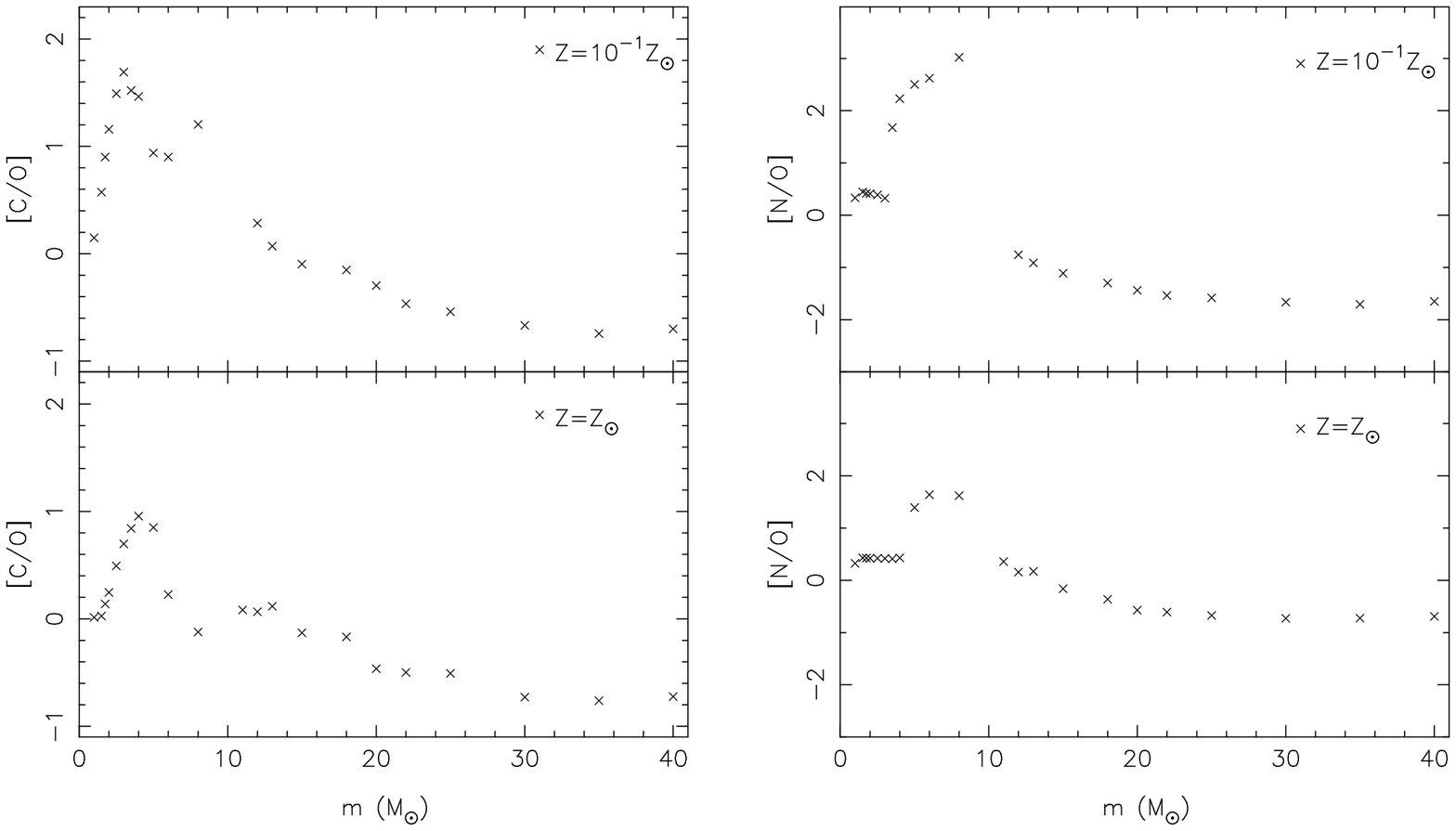}
\caption{Ratio of carbon (left panels) and nitrogen (right panels)
to oxygen for the adopted stellar yields - Woosley \& Weaver (1995) yields 
were adopted for Type~II supernova progenitors ($m$$\ge$10\,M$_\odot$), while
the low- and intermediate-mass yields of van~den~Hoek \& Groenewegen
(1997) were used for $m$$\le$8\,M$_\odot$.}
\end{figure*} 

The uncertainty in the stellar yields of zero metallicity stars was one of the
few caveats which accompanied the earlier semi-analytical models
of GM97 and Fields et~al. (2000).  As noted by
Marigo et~al. (2001) ``the distinct evolutionary behaviour of
these (zero metallicity) stars may also imply a very distinct
nucleosynthesis and chemical pollution of the interstellar medium''; 
at issue here is the production of carbon and nitrogen in Population~III
stars.  \it If \rm it could be shown that neither element was synthesised
or ejected into the ISM in the early Universe, a ``loophole'' could then exist
which would allow a WD-dominated Galactic halo without violating the
elemental abundance patterns in the halo today.  The new stellar evolution
models described by Abia et~al. (2001) suggest that large quantitative
differences do not exist between extant low-metallicity models (which do
undergo thermal pulses and eject significant carbon and nitrogen) and
Population~III (at least for masses 4$\leq$$m$/M$_\odot$$\leq$7; for
masses $m$$\le$4\,M$_\odot$, the Population~III yields remain 
uncertain).  Our models are qualitatively robust against the specific
choice of yield compilation.

\subsection{The Model}

We calculate semi-cosmological models, following the technique described
by Kawata (1999, 2001).  Our seed Galaxy is an isolated sphere upon
which small-scale density fluctuations corresponding to a Cold Dark Matter
(CDM) power spectrum are superimposed.  These initial density
fluctuations are generated using COSMICS (Bertschinger 1995).
The effects of longer wavelength fluctuations are incorporated by enhanced 
density of the sphere and the application of a spin parameter
$\lambda$ responsible for initiating rigid rotation of the overdense sphere. 
This spin parameter is defined
\begin{equation}
\lambda\equiv {J|E|^{1/2}\over GM_{tot}^{5/2}},
\end{equation}
\noindent
where $E$ is the total energy of the system, $J$ the total angular
momentum, $M_{tot}$ the total mass of the sphere (dark matter $+$
gas). Mass fluctuations (rms) in a sphere of radius $8 h^{-1}$\,Mpc are
denoted $\sigma_{8,in}$, and normalise the amplitude of the CDM power
spectrum. According to Katz \& Gunn (1992) and Steinmetz \& Muller (1995), a large value of $\sigma_{8,in}$ leads a large transfer of angular momentum from the gas particles to the dark matter particles, and a bulge larger than observed in the Milky Way; in order to avoid this, we consider the Galaxy to be formed in a an environment where the small scale density perturbation is smoother than the mean value suggested in standard CDM cosmology. We set $\sigma_{8,in}=0.1$. The expected collapse redshift is given by $z_c$. The amplitude
of the top-hat density perturbation $\delta_i$ is approximately related
(Padmanabham 1993) to initial and collapse redshift by
$z_c=0.36\delta_i(1+z_i)-1$. Thus $\delta_i$ is determined for a given
$z_c$ at $z_i$. We set $z_c=2.0$. The simulated
galaxies described here 
have total mass $M_{tot}=5\times 10^{11}$\,M$_\odot$, approximating
the total mass of the Milky Way within 50\,kpc (Wilkinson \& Evans 1999).  
For our models, the
initial conditions are uniquely determined by the choice of
$\lambda$, $M_{tot}$, $\sigma_{8,in}$, and $z_c$. 

Using a comparable semi-cosmological model, Katz \& Gunn (1991) found that a 
seed galaxy with a large spin parameter evolves into a disk
system.  Our best Milky Way models require $\lambda=0.07$, in order to
remain consistent with the Galaxy's present-day disk properties.
We assume a flat ($\Omega_o=1$) universe with baryon fraction
$\Omega_b=0.1$ and Hubble constant H$_o=50$\,km\,$s^{-1}$\,$Mpc^{-1}$; evolution within isolated galaxy simulations using top hat overdensities are not greatly influenced by choice of cosmology. Evolution is traced from an initial redshift $z_i=40$ to the present.

The initial conditions used are somewhat artificial, and will not exactly reflect hierarchical formation of the Milky Way as expected in CDM cosmologies. However, this approach still traces hierarchical build up of a galaxy, self consistently treating its chemical and dynamical history, and is  sufficient for the purposes of  documenting the effects between different IMFs.  

Our study relies on the chemical enrichment processes of SPH, as described in section 2.1.3. In order to ensure that such processes are not determined by the resolution adopted, we perform the study at three different resolution regimes. The resolution regimes employ 6142, 18342, and 40958 particles (initially split evenly between gas and dark matter). The results presented in this paper are from the highest resolution models; the results of different resolution regimes varied only marginally quantitatively, and the analysis and conclusions presented hold in all three; we therefore only present the highest resolution models. Our highest resolution models have  dark matter (baryonic) particles of mass $2.2\times10^{6}$ M$_\odot$ ($2.4\times 10^{5}$ M$_\odot$), which is comparable in resolution with other recent simulations of disk galaxy formation (e.g. Abadi et. al 2003)\footnote{In galaxy formation simulations including cooling, the physical resolution determines the computational costs rather than the total number of particles.}. The gravitational softening length is 0.79 kpc (1.6 kpc) for baryonic (dark matter) particles.

As noted earlier, the primary ``variable'' in our simulations is the
initial mass function (IMF).  Three different scenarios have been
considered here - models 1 and 2 assume the Salpeter (1955) and
Chabrier et~al. (1996) IMFs, respectively, regardless of metallicity;
model 3 assumes the Chabrier et~al. IMF (wdIM)  for metallicities below
5\% solar, and the Salpeter IMF (sIMF)  elsewhere, as
summarised in Table~3.

\begin{table}\label{models}
\caption{The three models adopted in the current analysis; the initial
mass function (IMF) is the only variable.}
\begin{tabular}{ccc}
         & \multicolumn{2}{c}{IMF}                    \\
   Model & $Z<0.05\; Z_\odot$ & $Z>0.05\; Z_\odot$    \\
 model 1 & Salpeter           & Salpeter              \\
 model 2 & Chabrier           & Chabrier  	      \\
 model 3 & Chabrier           & Salpeter              \\  
\end{tabular}
\end{table}

\section{Results}
Figures~3 and 4 show the X-Y and X-Z projections of
the morphological evolution for model 3, over the redshift range
$z=2.7$ to $z=0.9$; models 1 and 2 are qualitatively
similar.  The Z-axis corresponds to the initial rotation axis.  Sub-clumps 
associated with early star formation can be traced to initial small-scale
density fluctuations, and are seen by redshift $z\approx 2.5$.  These sub-clumps
merge hierarchically until redshift $z\approx 1.7$; by redshift
$z\approx 1.6$, large sub-clumps merging in the centre of the Galaxy (where the disk is 
forming) is responsible for the bulk of star formation.    An epoch of accretion (1.2$<$$z$$<$0.9) sees a majority of the
high-metallicity ($Z > 0.05 Z_\odot$) stars formed, and the disk-like
structure of the Galaxy becomes evident. The star formation
history of model~3 is shown in Figure~5. The system evolves little 
morphologically after redshift $z\approx 0.8$.
While not shown, we confirm that the radial scale lengths for our 
simulated disk at $z = 0$ (both gas and stellar) match those of the
Milky Way.

\hspace{-2cm}
\begin{figure*}

\plotone{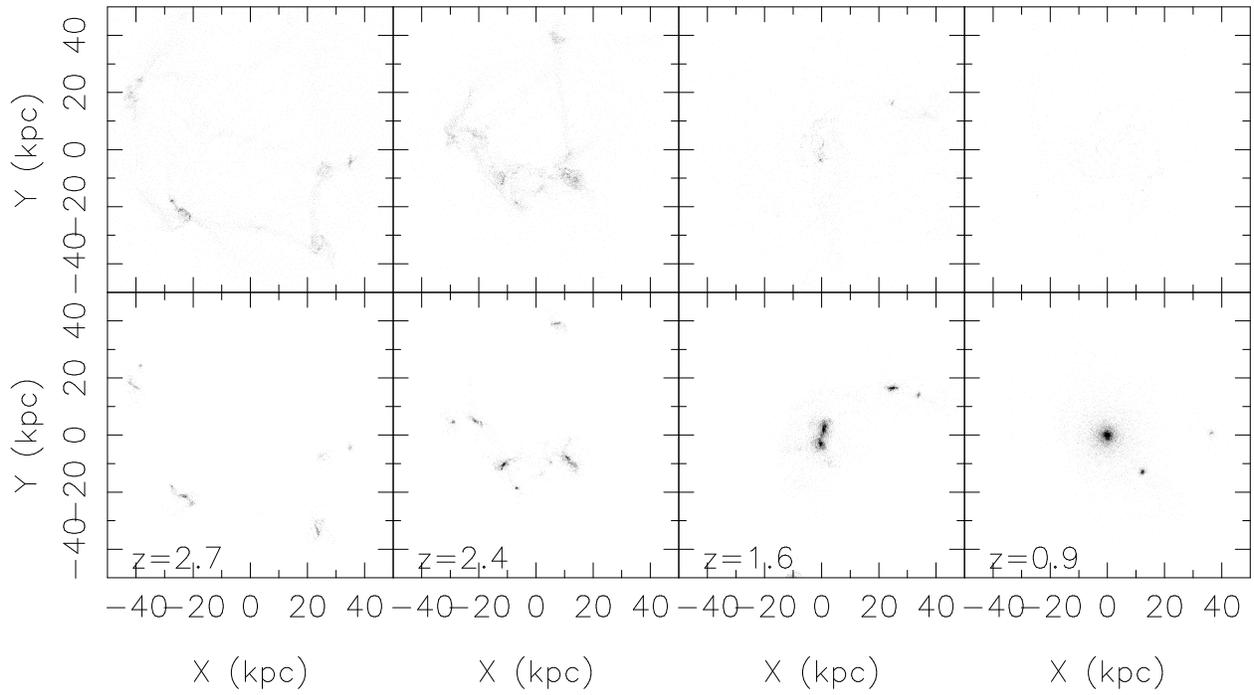}
\caption{Projection onto the X-Y plane of the evolution of gas (upper
row) and stars (bottom row)
during the major star forming epoch of model~3.}
\label{fig3}
\end{figure*} 

\begin{figure*}
\plotone{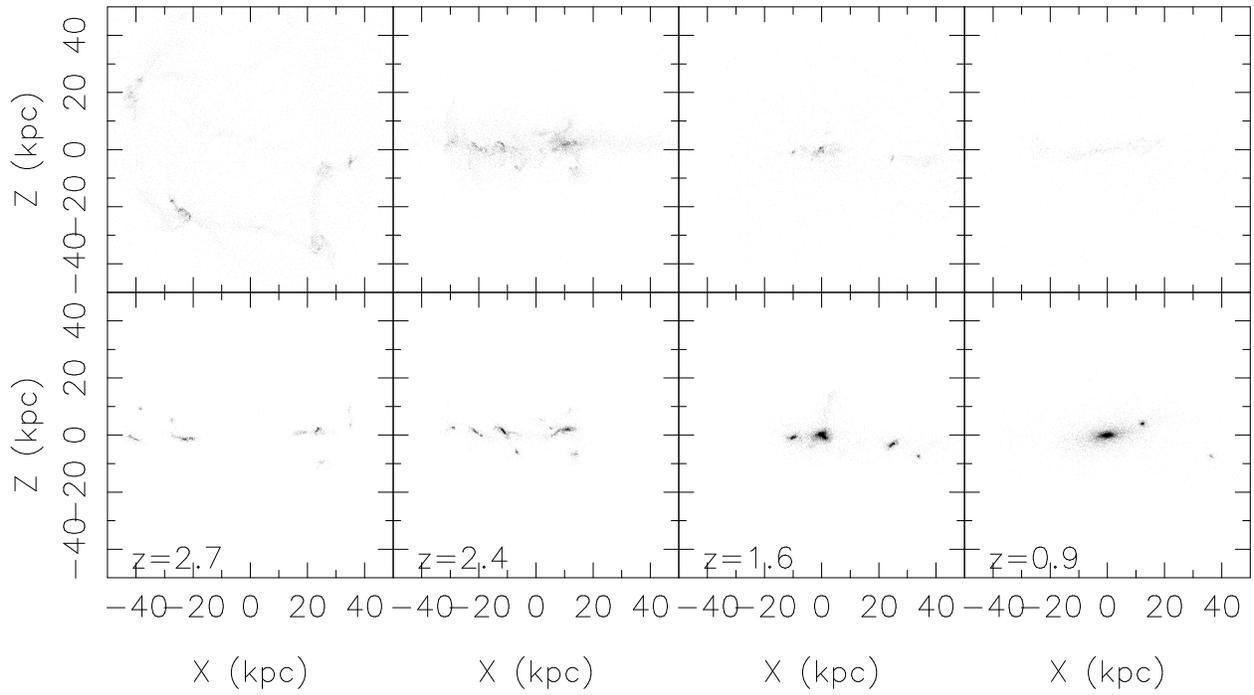}
\caption{Projection onto the X-Z plane of the evolution of gas (upper
row) and stars (bottom row) 
during the major star forming epoch of model~3.}
\label{fig4}
\end{figure*} 

\begin{figure*}
\plotone{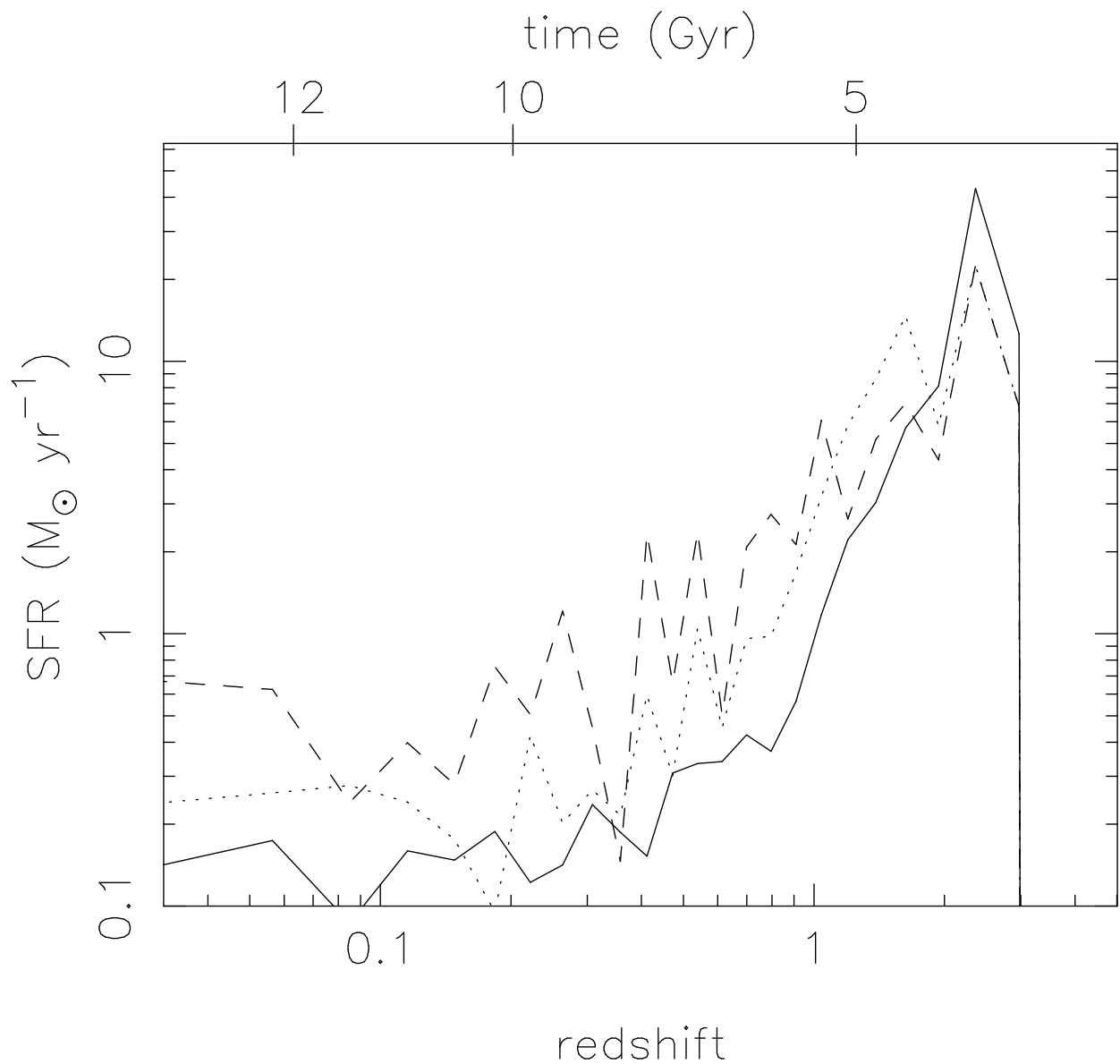}
\caption{Evolution of the global star formation rate (SFR) of model~1 (solid line), model~2 (dashed) and model~3 (dotted).}
\label{fig5}
\end{figure*}

\begin{figure*}
\plotone{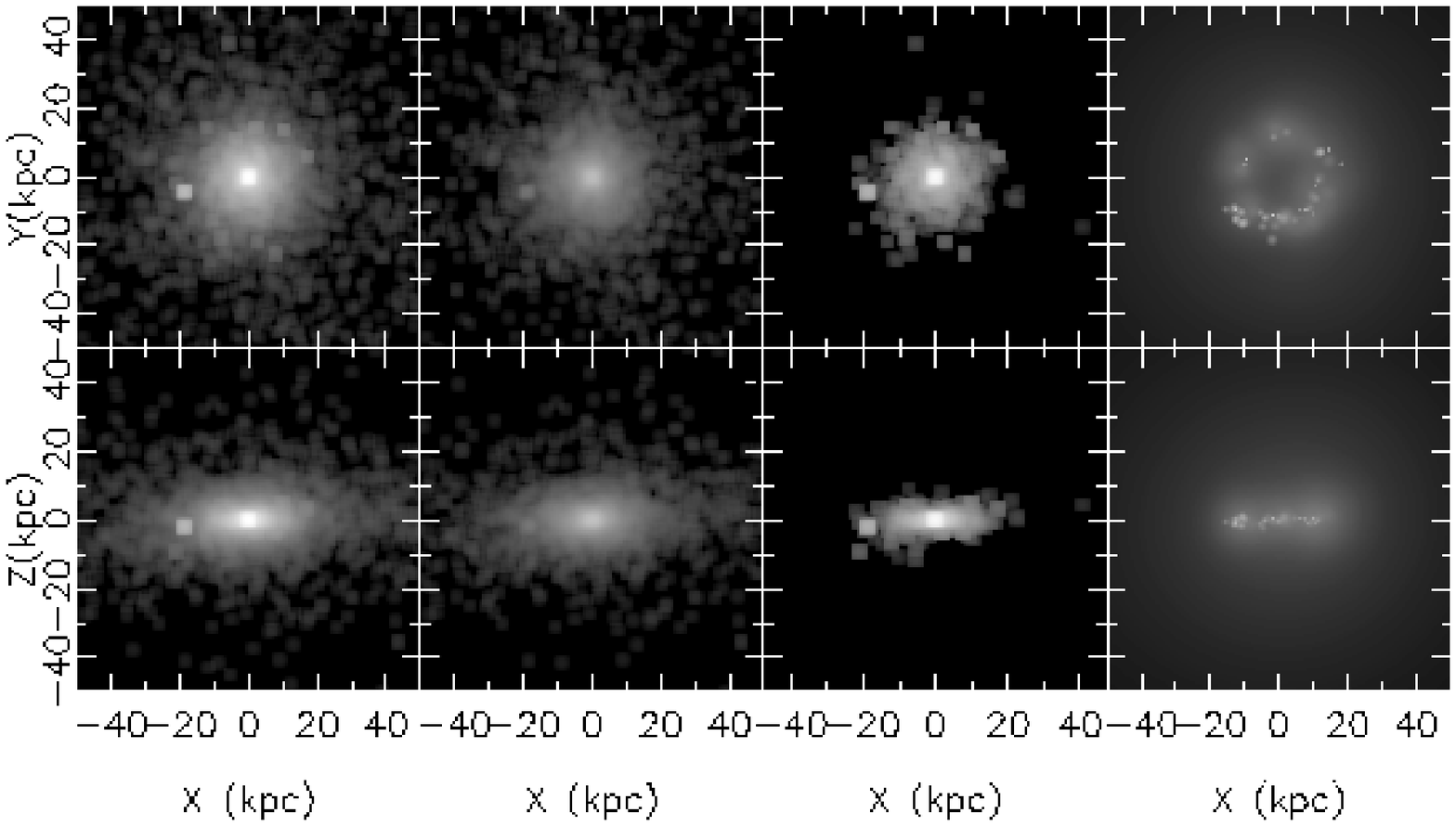}
\caption{Density map of the final ($z=0$) distribution of star particles (left panel), low-metallicity
($Z<0.05 Z_\odot$) stars, high-metallicity ($Z>
0.05 Z_\odot$) stars, and gas particles, for model~3, in the X-Y (upper row) 
and X-Z (bottom row) planes.}
\label{fig6}

\end{figure*}

Density maps of the  final distributions ($z=0$)
of star and gas particles for model~3 are shown in
Figure~6. The X-Y (upper row) and X-Z (bottom row)
planes are shown- the low-metallicity ($Z<0.05 Z_{\odot}$) stars of the Galaxy preferentially populate the halo and thick disk, while the
metal-rich ($Z>0.05 Z_{\odot}$) component populates the bulge and thin disk.

\subsection{White Dwarf dominated Halo}
In this section we investigate whether our models~2 and 3  have resulted in a halo which has contains a significant amount of mass tied up in white dwarf stars.
We make a comparison of the resulting
halo and disk white dwarf (WD) densities of our three models, derived using the different assumed 
IMFs. The halo stars are defined as star particles in the region $4<R_{XYZ}<20$ kpc and $|Z|>4$ kpc. 
The {\it mean} halo 
white dwarf  number densities $n$, for each of the IMF models discussed 
in \S~2 are shown in Table~4, where $n$ is defined as

\begin{equation}{ n={1\over V}\sum_im_{s,i} \int_{1 M_\odot}^{8 M_\odot} \Phi(m)\,{\rm d}m }\end{equation} 

The summation is over star particles which are older than 1 Gyr (the lifetime of a 2 M$_\odot$ stars, and also the time delay adopted in our chemical evolution modeling of white dwarf progenitors- recall Section 2.1.3) and found in the above region for the halo. It is important to note that the absolute values of the densities is sensitive to the definition of the region selected; 
the densities are shown to provide details of how the number densities of white dwarfs altered between the three models.

As expected, model~2 (Chabrier IMF at all metallicities) results in
significantly more white dwarfs 
in the halo (by a factor of 3.8), 
when compared with 
model~1 (Salpeter IMF at all metallicities).
The variable IMF of model~3 (Chabrier at low-metallicities and
Salpeter elsewhere) results in a factor of 3.5 increase in the
halo white dwarf population with respect to model~1. In other words, model~3 increases the number of halo white dwarfs to almost the same degree as model~2, when compared with the canonical Galactic
models which use a Salpeter IMF alone (i.e., model~1).  The ability
of model~3 to increase the  number of halo white dwarfs by a similar amount to model~2, makes it an important
component of chemo-dynamical models aimed at maximising the number of halo white dwarfs, while allowing higher metallicity stars to be distributed according to the Salpeter IMF.

We are further able to examine the ratio of white dwarf stars to main sequence stars in the halos of the three models.  Using appropriate IMFs, we calculate 
$ \sum_{i}^{halo} {n_{{\rm WD},i}/ n_{ms,i}}$ giving  a number ratio of white dwarf to  main sequence stars in the halo region of 0.19 for model~1, 1.2 for model~2 and 1.0 for model~3. Here the number of white dwarfs is
$n_{{\rm WD},i}=\int_{1 M{_\odot}}^{8 M{_\odot}}\Phi (m) {\rm d}m\; {\rm if \; t}_{age,i}> 1$ Gyr and the number of main sequence stars is $ n_{ms,i} = \int_{0.3 M{_\odot}}^{1 M{_\odot}}\Phi (m) {\rm d}m$. We see that model~3 significantly increases the number of white dwarf halo stars, as required by our white dwarf halo scenario.
This highlights that models~2 and 3 have resulted in creating a stellar halo which has a significant mass in the form of white dwarf stars. We found that this result was robust over a range of ``regions'' which could be used to define the halo.  Unlike the values of densities quoted in Table~4, {\it this allows a comparison which is not directly sensitive to the region chosen as defining the halo}.

The differences in evolution and final morphology of models~1 and 2 from model~3 are not significant enough to show as separate diagrams in \S~3, yet they do have some impact on the total stellar densities of our simulated halo region. Factors which will influence these differences are; the different ejected mass due to different IMFs result in different masses, and hence dynamics, of the remnant stellar particles; different ejecta also results in different gas enrichment and mass and different star formation rates (Figure~5); different cooling rates due to differences in metallicity of enriched gas particles; different energy feedback due to different supernova rates (although our implementation of such  feedback is known to be highly inefficient in Tree-SPH codes). These are ``secondary'' effects of a different IMF on the white dwarf densities quoted in this section. As it is difficult to disentangle these effects, Table~4 shows the total halo stellar densities of the three models in order to indicate the magnitude of their combined effects. Here, total stellar density means the value of the summation of the mass of star particles in the halo region, divided by the volume of the halo region. There is little difference between the models in such total stellar densities and thus these secondary effects are not what drives the differences in white dwarf densities between the models. The difference in the number densities of white dwarfs between models can therefore be attributed  to the variation of the adopted IMFs.

\begin{table*}\label{densities}
 \begin{minipage}{140mm}
 \centering
  \caption{Total stellar and white dwarf number densities (in units of pc$^{-3}$) for the three 
models described here.}
  \begin{tabular}{@{}cccccc@{}}
    & model 1 & model 2& model 3 \\
         &           &  &                           \\
Halo total stellar density & $1.4 \times10^{-4}$        &    $ 1.5\times10^{-4}$ &   $1.5 \times10^{-4}$     \\
 Halo white dwarf density & $1.7\  \times10^{-5}$  &    $ 6.0  \times10^{-5}$ &   $6.5  \times10^{-5}$   \\
\end{tabular}
\end{minipage}
\end{table*}

\subsection{Chemical Constraints}

In this section, we examine the constraints set by the observartions of chemicals, in particular carbon, nitrogen and oxygen, set on the viability of our three models. We are especially interested in  whether the variable IMF of model~3 alleviates the
problem of carbon and nitrogen overproduction relative to oxygen,
as originally pointed out by GM97.  Figure~7 shows the [C/O] and [N/O] 
abundance ratios
as a function of global metallicity [Z]\footnote{[Z]=log$_{10}$(Z/Z$_{\odot}$)} for the stellar particles in the solar neighbourhood for all models, to compare with observations of solar neighbourhood stars. We define the solar nighbourhood as the region  bounded by 4$<R_{\rm XY}<$10~kpc 
and $|{\rm Z}|<$1.5~kpc. Carbon and nitrogen are highlighted
due to the fact that their dominant nucleosynthesis sites are 
the intermediate mass white dwarf precursors under discussion here.

\begin{figure*}

\plotone{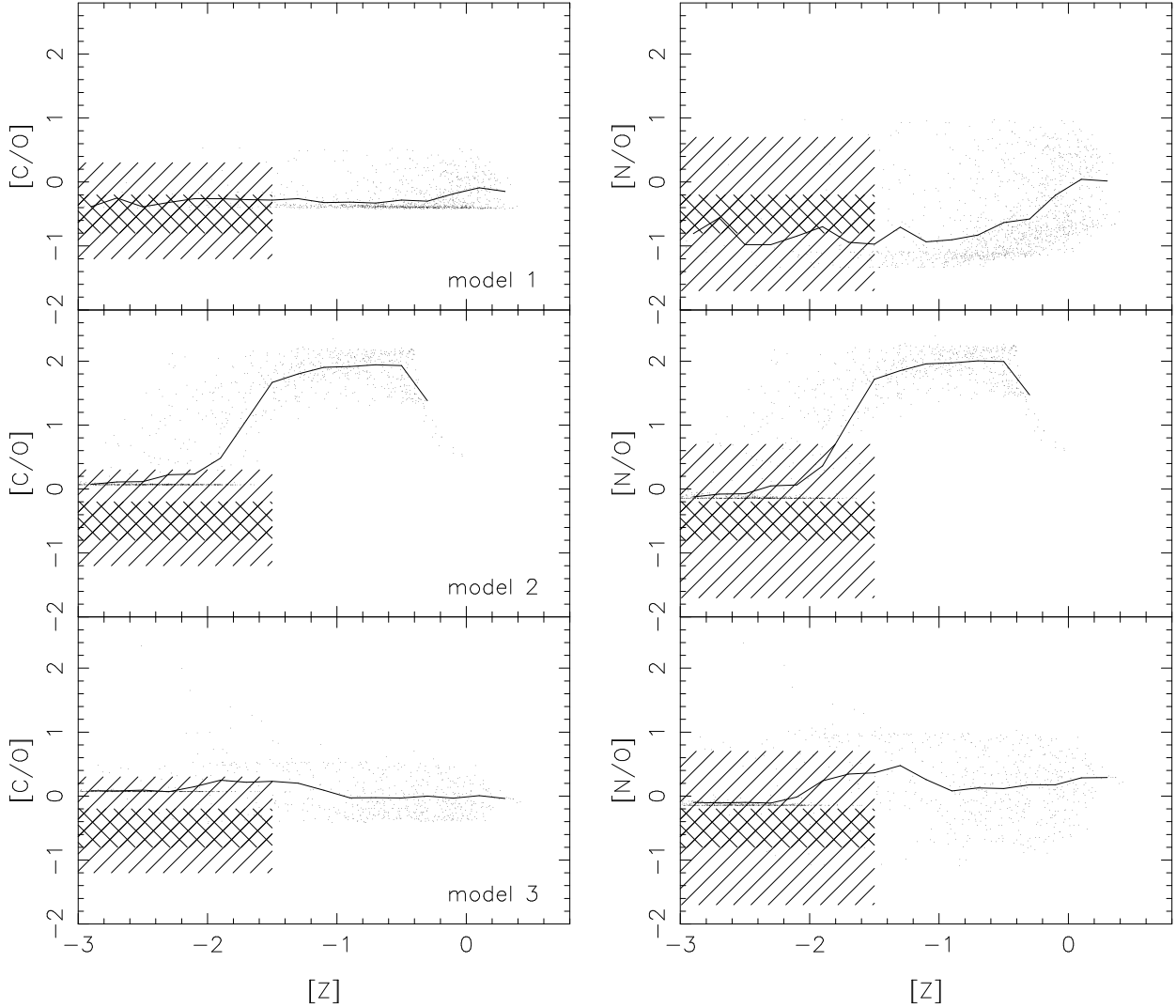}
\caption{The values of [C,N/O] for simulated star particles in the
solar neighbourhood plotted against
their metallicity [Z] at redshift $z$=0. 
The results of model~1 are in the upper panels,
below which are the results for models~2 and 3, respectively. 
The trend lines show the
mass-weighted mean values of the simulations. The
hatched regions correspond to the observational
constraints listed by Timmes et~al. (1995). All of the $\sim$150 halo 
stars shown in
Figures~13 and 14 of Timmes et al. lie within the bounds of the
outer hatched regions, while $\sim$80\% of the sample lies within the
inner regions.}
\label{fig7}
\end{figure*}

We constrain our models with the observed abundance distribution pattern
of carbon, nitrogen, and oxygen seen in the metal poor Population~II stars, which are dominated by halo stars (hashed regions in each panel of Figure~7). We remind the
reader that Type~Ia supernovae have not been included in the 
present anaylysis.  As such, we take the total metallicity [Z] to act as a 
proxy for [Fe/H], without violating the validity of the simulations (since
Type~Ia supernovae are not critical contributors to the Galaxy's
CNO abundances).  The points in Figure~7 correspond to those simulated
star particles which reside, at the present-day, in the solar neighbourhood defined above - no kinematical (i.e., disk versus halo
kinematics) cut was applied as the observational data we compare to, is of solar neighbourhood stars without kinematic selection.

The solid lines in each panel of Figure~7 correspond to the mass-weighted
mean stellar metallicity of the simulation.  For model~1, we can see that
there is essentially no variation in [C,N/O] for halo metallicities
(i.e., [Z]$<$$-$1), with the simulated halo particles occupying the
same parameter space as the data (hashed regions).  As the chemical
enrichment from low- and intermediate- mass stars begins to become
important, [C,N/O] approaches the solar ratio and the dispersion
increases, both being in keeping with the observations of the solar
neighbourhood distributions (Timmes et~al. 1995).  Not surprisingly, 
this is consistent with the conventional picture in which the sIMF (model~1)
scenario is assumed to hold.

Model~2 (middle panels of Figure~7) shows the clear signature of carbon 
and nitrogen overproduction.  As the intermediate mass stars responsible
for the nucleosynthesis of (the bulk of) these elements pass through the
asymptotic giant branch, the resulting [C/O] and [N/O] quickly exceeds the
halo observational constraints by more than two orders of magnitude -
an invariant wdIMF model cannot be made consistent with the data (this is consistent with the analytical chemical evolution analysis of GM97).

The lower panels of Figure~7 demonstrate that model~3 (the variable IMF) does
mitigate (somewhat) the overproduction of carbon and nitrogen at low
metallicities.  Below [Z]$\approx$$-$1, model~3 traces model~2; above
[Z]$\approx$$-$1, the Salpeter (1955) component of the variable IMF
begins to dominate star formation (and its associated element production),
suppressing the extreme overproduction inherent in the invariant wdIMF
model~2.  Ultimately, [C/O] and [N/O] approach that of the model~1 (sIMF),
reaching the scaled-solar ratio at [Z]$\sim$0.  While a definite
improvement (in the sense that the factor of $>$100 overproduction
in carbon and nitrogen seen in model~2 no longer exists), 
model~3 still leads to a factor of $\sim$5-10 carbon and nitrogen
excess (with respect to oxygen) at metallicities in the range
$-$2.5$<$[Z]$<$$-$1.

A couple of features of the plots in Figure~7 are worth further comment here. In models 2 and 3 we notice a number of stars in a horizontal line with values of [C,N/O]$\sim$ 0. These stars were formed from gas polluted by massive stars, prior to intermediate star's yields being ejected. Reference Table~1 which shows that the wdIMF massive star's ($> 10$M$_\odot$) integrated yields have [C,N/O]$\sim$ solar. Also note the limit of [C,N/O] of $\sim 2.2$ in all models in Figure~7 which comes from the low metalicity intermediate mass stellar  yields, as seen in Table~2.

\section{Discussion}

Ascribing the detection of the 13-17 microlensing events seen toward the LMC 
to an intervening halo white dwarf (WD) population implies that perhaps
20\%$-$30\% of Galaxy's dark matter is tied up in these stellar remnants
(Alcock et~al. 2000).  This would correspond to a local WD halo number 
density of $\sim$2$\times$10$^{-3}$\,pc$^{-3}$.  In contrast, Gould et~al.
(1998) concluded, using subdwarf star counts and an assumed Salpeter (1955)
IMF, that the local spheroid WD number density was 2.2$\times$10$^{-5}$\,pc$^{-3}$
(i.e., two orders of magnitude smaller).  More recently, the high proper motion
survey for cool white dwarfs conducted by Oppenheimer et~al. (2001)
claimed a local halo WD number density of 
2.2$\times$10$^{-4}$\,pc$^{-3}$.\footnote{The re-analysis conducted by
Gibson \& Flynn (2001) suggest that the Oppenheimer et~al. (2001) 
number density should be revised downward by 40\% to 
1.3$\times$10$^{-4}$\,pc$^{-3}$.}  Whether the Oppenheimer et~al. WDs
are actually members of the halo, as opposed to the thick disk, is
still a matter of debate (e.g. Gyuk \& Gates 1999; Reid \& Sahu 2001),
although Hansen (2001) argues that if the halo is dominated by WDs, 
$\sim$80\% of these remnants lie beyond the detection limit of 
Oppenheimer et~al. - i.e., the WD-dominated halo issue has not yet
been resolved and future (deeper) surveys must be undertaken before
the final word on the subject can be made.

Regardless of the current observational constraints, it is recognised that
populating the Galactic halo with a dynamically-dominant WD component 
demands a significant modification to the underlying stellar IMF
(e.g. GM97).
A Chabrier (1996)-style IMF, peaked in the WD progenitor range, is one
such possibility, although this IMF cannot apply at all times since
we observe today a fairly conventional Salpeter (1955) or Scalo
(1986)-like distribution.  Metallicity may be the controlling factor in
``switching'' from the Chabrier-like to Salpeter-like IMF; a physical
motivation for such a switch can be found in the studies of Yoshii \& Saio (1987),
Nakamura \& Umemura (2001), Bromm et~al. (2001) and Abel, Bryan \& Norman (1999, 2001) . Making our IMF dependent upon metallicity in our variable model is further motivated by the different MDF's of the Galactic disk and halo regions; if one wishes to populate only the halo with large numbers of white dwarf stars, then it is the low metalicity stars which should have an IMF biased toward white dwarf progenitor stars. This appears necessary in order to explain MACHO observations by invoking white dwarf stars.

Our chemodynamical simulations employing a hybrid IMF (WD-dominated at
metallicities [Z]$<$$-$1.3, and ``conventional'' at metallicities
[Z]$>$$-$1.3) are successful in increasing the Galactic halo WD fraction
(in comparison with models employing a Salpeter 1955 IMF alone)
without violating present-day observable local carbon and nitrogen Pop II abundance 
constraints.
These results are caused by low  metallicity stars preferentially being found in the halo region as seen in Figure~6. Low metallicity stars are formed before and during the major star formation epoch caused by the collapse of the whole system. These stars are scattered by violent relaxation. Later, gas polluted by past star formation accretes smoothly and the metal rich disk component developes. This picture is consistent with previous chemo-dynamical studies of a Milky Way like galaxy (Steinmetz \& Muller 1995; White et al. 2000; Bekki \& Chiba 2001; Abadi et~al. 2003). This metallicity segregation is what allows our simple variable IMF to populate the low metallicity stars of the halo with a large number of white dwarf stars. The discrepency between observed metallicity distributions of Galactic halo and disk stars mean that our scheme of varying the IMF according to metallicity would seem appropriate, indeed necessary, if we are to explain MACHO events by white dwarfs.

That said, our best models were only capable of increasing the number of halo white dwarfs by a factor of 3-4 (Table~4) before violating the constraints of an overproduction of carbon and nitrogen compared to oxygen. This factor is between our simulated models, and unlike the absolute densities  is not significantly affected by the region selected as defining the halo. How does this factor relate to observed white dwarf densities?  The Gould et~al. value of 2.2$\times 10^{-5}$ pc$^{-3}$ assumes a Salpeter IMF; increasing this by a factor of 3-4 corresponds to halo WD mass fractions in of $\sim$1-2\% -  halo WD mass fractions beyond this 
are not supported by our simulations.

Attempts to increase the halo WD fraction further necessitate significant
revision to our fairly conventional star formation formalism.  We attempted
to force the halo WD issue by increasing substantially the efficiency
of star formation
($c_*$ in equation~\ref{sfr}) in low-metallicity star-forming
regions. We also tried decreasing the density threshold for star
formation $\rho_{crit}$ in these same regions.\footnote{both these simulations were done using n$_{tot}=18342$}.  Both \it ad hoc \rm
revisions introduced moderate increases in the halo WD fraction.  For
example, increasing $c_*$ by a factor of ten and decreasing
$\rho_{crit}$ by the same factor of ten, led to
a halo WD density of 1.3$\times$10$^{-4}$\,pc$^{-3}$ (c.f. model~3
entries of Table~4).  While such a
density would be consistent a factor of $\sim 8$ incease of halo white dwarf densities over the Salpeter IMF model, this
``extreme'' model suffers from the same carbon and nitrogen
overproduction which plagued model~2 (Section~3), mitigating its
usefulness.

The hybrid IMF preferred here
varies in a very simple manner and that the 0.05\,Z$_\odot$ transition
metallicity for changing from the wdIMF to sIMF regime was a somewhat 
arbitrary choice.  We found that  adopting a lower metallicity for the transition
(e.g. 0.01\,Z$_\odot$) does diminish the carbon and nitrogen overproduction,
but at the expense of reducing the halo WD mass fraction. 
 Thus, it becomes clear that a trade off exists between increased white dwarf halo densities and the overproduction of carbon and nitrogen compared to oxygen; any significant increases in white dwarf numbers are accompanied by worsening of the overproduction of carbon and nitrogen, as well as contamination of the disk with an unacceptable number of white dwarf stars. Our simulations suggest that attempts at increasing the number of  halo white dwarf beyond that obtained in model~2    ($\sim$1-2\% of the halo, by mass) will violate these constraints.

Our best hybrid IMF model, while an improvement over purely WD-dominated
IMF models, still overproduces stars with supersolar
[C/O] and [N/O] ratios in the metallicity range $-2.5 <$[Z]$<$$-$1.
Modifications to the star formation efficiency and threshold were incapable
of eliminating this overproduction.  As such, more extreme scenarios may
need to be invoked in order to suppress the formation of these stars.
An early Galactic wind, fine-tuned such that the carbon and nitrogen
ejecta from the wdIMF component of our hybrid IMF is removed from the
system before being incorporated into the stars which ultimately
form the Population~II halo of the Milky Way, is one such potential
modification (e.g. Fields et~al. 1997; GM97).
Although we do take into account the energy
feedback of SNe (Section~2.1.3), our favoured model~3 does not develop
a Galactic wind.  That said, the exact efficiency of SNe feedback
is uncertain; in a future work, we will explore more extreme feedback
scenarios, in an attempt to suppress excess carbon and nitrogen 
overproduction at early times.  At that time, we will also include
the effects of Type~Ia SNe, although we remind the reader that this
inclusion has little impact upon the results shown here (as SNe~Ia
are negligible contributors to the carbon, nitrogen, and oxygen
yields discussed here - see Section~2.1.3).

\section{Summary}

We have undertaken Galactic chemodynamical simulations of the formation and
evolution of the Milky Way, exploring a region of parameter space designed
to test the hypothesis that the Galaxy's halo has a significant dynamical
white dwarf component (perhaps as high as 30\% by mass).  This is the
first numerical examination of the WD-dominated hypothesis, and a natural
extension to the earlier semi-analytical investigation of Gibson \&
Mould (1997).  Consistent with Gibson \& Mould, we have found that
carbon and nitrogen overproduction (by factors in excess of 100)
are endemic to purely WD-dominated IMF models.  For the first time, we have
shown that a hybrid WD-dominated/Salpeter style IMF with a transition
metallicity between the two of Z$=0.05\,$Z$_\odot$ can lead to a 
Milky Way model which is marginally consistent with virtually all
available present-day constraints.  A mild overproduction of carbon
and nitrogen remains (although this overproduction is reduced by more
than an order of magnitude); an early Galactic wind may alleviate this
overproduction, an hypothesis we will explore in a future study.
Our best model can lead to a halo WD mass fraction a
factor of $\sim 4$ greater than expected with a more conventional
IMF (corresponding to $\sim 1-2\%$ of the mass of the halo), but values in excess of this cannot be accommodated within
our chemodynamical framework.

\section*{Acknowledgments}
We acknowledge the support of the Australian Research Council through its
Large Research Grant Program (A0010517).  CB acknowledges receipt of
an Australian Postgraduate Award.  We thank Chris Flynn for his helpful
advice during the completion of this manuscript.  
The simulations described here
were performed at the Victorian Partnership for Advanced Computing and
the Beowulf Cluster at the Swinburne University Centre for Astrophysics
\& Supercomputing.

\end{document}